\documentclass[preprint,12pt,authoryear]{elsarticle}

\usepackage{amssymb}
\usepackage{amsmath}
\usepackage{amsthm}
\usepackage{lineno}
\usepackage{multirow}
\usepackage[table]{xcolor} 
\usepackage{colortbl} 
\usepackage{url}
\usepackage[bottom=3cm,footskip=35pt]{geometry}

\journal{Ocean Modelling}

\newcommand{\NetName}{\emph{InWaveSR}} 
\newcommand{\BlockName}{\emph{HF-ResBlock}} 
\begin{document}

\begin{frontmatter}

\title{InWaveSR: Topography-Aware Super-Resolution Network for Internal Solitary Waves}

\author[1]{Xinjie Wang}
\author[1]{Zhongrui Li}
\author[1]{Peng Han}
\author[2]{Chunxin Yuan\corref{yuan}}
\cortext[yuan]{Corresponding author}
\ead{yuanchunxin@ouc.edu.cn}

\author[3]{Jiexin Xu}
\author[1]{Zhiqiang Wei}
\author[1]{Jie Nie}
\affiliation[1]{organization={Department of Computer Science and Technology, Ocean University of China},
                     city={Qingdao},
                     postcode={266100},
                     country={China}}
\affiliation[2]{organization={School of Mathematical Science, Ocean University of China},
                     city={Qingdao},
                     postcode={266100},
                     country={China}}
\affiliation[3]{organization={State Key Laboratory of Tropical Oceanography, South China Sea Institute of Oceanology, Chinese Academy of Sciences},
                     city={Guangzhou},
                     postcode={510301},
                     country={China}}

\begin{abstract}
The effective utilization of observational data is frequently hindered by insufficient resolution. To address this problem, we present a new spatio-temporal super-resolution (STSR) model, called \NetName. It is built on a deep learning framework with physical restrictions and can efficiently generate high-resolution data from low-resolution input, especially for data featuring internal solitary waves (ISWs). To increase generality and interpretation, the model \NetName~uses the primitive Navier-Stokes equations as the constraint, ensuring that the output results are physically consistent. In addition, the proposed model incorporates an \BlockName~component that combines the attention mechanism and the Fast Fourier Transform (FFT) method to improve the performance of the model in capturing high-frequency characteristics. Simultaneously, in order to enhance the adaptability of the model to complicated bottom topography, an edge sampling and numerical pre-processing method are carried out to optimize the training process.
On evaluations using the \textit{in-situ} observational ISW data, the proposed \NetName~achieved a peak signal-to-noise ratio (PSNR) score of $36.2$, higher than those of the traditional interpolation method and the previous neural network. 
This highlights its significant superiority over traditional methods, demonstrating its excellent performance and reliability in high-resolution ISW reconstruction.
\end{abstract}



\end{frontmatter}

\section{Introduction}\label{sec: Introduction}
Internal solitary waves (ISWs) are a unique class of nonlinear internal waves~\citep{guo2014}, ubiquitous in the world's oceans, see the review by \citet{mixing2020}.
These waves play a crucial role in modulating oceanic processes, such as sediment resuspension~\citep{deepwell2020particle}, submarine sand dunes formation~\citep{ma2016footprints}, marine ecosystem regulation~\citep{woodson2018the} and climate system \citep{mixing2020}.
In internal wave dynamics, the amplitude of ISWs serves as a crucial parameter to evaluate their energy, while also playing a decisive role in their interactions with ocean currents and marine ecosystems~\citep{meng2024energy}. 

Due to the variations of background currents, bottom topography, and the inclusion of the Earth's rotation, ISWs usually possess trailing waves, which is a significant characteristic of ISWs and is critical in governing energy dissipation and dispersion mechanisms~\citep{zhang2018}.
Understanding and observing the amplitude of ISWs, along with their associated trailing waveforms, are essential to uncover their underlying dynamics~\citep{la2024on}.
However, the dynamics of ISWs often exhibit pronounced nonlinearity and complexity due to the significant influence of intricate bottom topography~\citep{rattray1960coastal}. 
This complexity increases the difficulty of their dynamical analysis and imposes higher demands on the training datasets for models, thereby affecting their performance~\citep{wang2020evolution}.

To gain deep insights into the mechanisms of ISWs, high-precision, and high-resolution spatio-temporal data is crucial, nevertheless, the observational data usually do not satisfy the sufficient spatial and temporal coverage. 
Thus, numerical models based on the primitive Navier-Stokes equations become an important tool.  
Traditionally, numerical simulation methods relied on techniques grounded in physics-constrained frameworks to achieve high-resolution data while ensuring physical consistency.
By strictly adhering to physical laws, such as mass conservation, momentum conservation, and energy conservation, these methods have played a critical role in generating high-precision ocean dynamical data~\citep{marshall1997finite}.
The finite difference method and the finite element method are typical examples, and they utilize highly refined computational grids that strictly adhere to physical laws, yielding highly accurate simulation results~\citep{brenner2008mathematical}.
However, the main drawbacks of these methods are that they are computationally intensive, and present significant challenges for large-scale, long-term simulations.

In recent years, to address the limitations of traditional methods, researchers have started incorporating deep learning technology into numerical simulation.
Deep learning has been shown to effectively reduce computational costs while maintaining reasonable accuracy in various domains, including fluid dynamics and environmental modeling~\citep{raissi2019physics, brunton2020machine}. 
Building on the advances in deep learning, researchers have proposed combining deep learning networks with physical constraints to further enhance computational efficiency and ensure adherence to fundamental physical laws~\citep{liu2021dual, wang2022transflownet}. 
The core concept lies in embedding physical constraints into data-driven models, leveraging the efficient learning capabilities of neural networks to capture complex flow field characteristics while ensuring that the generated results adhere to dynamical consistency guided by physical laws.
These hybrid approaches integrate low-resolution simulations with data-driven techniques, striking a balance between computational efficiency and physical consistency~\citep{esmaeilzadeh2020meshfreeflownet}.
Most existing studies focus on small-scale or idealized flow fields, but their applicability to complex, real-world oceanic phenomena remains limited~\citep{chen2020geom}.
Especially, ISWs, a typical nonlinear oceanic phenomenon, involve strong nonlinear convection terms and pronounced high-frequency characteristics~\citep{grimshaw2010internal, xue2013iswamp}. 
These features present substantial challenges for numerical modeling and data-driven methods, as accurately capturing ISW dynamics requires both high spatio-temporal resolution and strict physical consistency~\citep{dong2016iswsat, song2021observations}.
Furthermore, complex bottom topography significantly affects the dynamical behavior of ISWs \citep{rattray1960coastal}, complicating the preparation of training datasets, thereby adversely impacting the model's performance~\citep{wang2020evolution}.

In this study, a novel spatio-temporal super-resolution (STSR) flow field reconstruction network, referred to as \NetName, is proposed, which combines the strengths of physical modeling and deep learning.
The training dataset for the network was generated by the Massachusetts Institute of Technology general circulation model (MITgcm)~\citep{dor2023, jia2024FMS}, which is known for its flexibility and accuracy in simulating diverse ocean conditions.
Based on fully non-linear and non-hydrostatic Navier-Stokes equations, MITgcm generates high-quality ISW datasets that provide a basis for training \NetName.
On that basis, \NetName~employs an encoder module that uses the Fast Fourier Transform (FFT) method ~\citep{chi2020fast}, attention mechanisms and residual structures to effectively extract the high-frequency and complex features of ISWs.
The use of FFT-based methods~\citep{yamanaka2017fast}, along with residual structures~\citep{he2016deep} and attention mechanisms~\citep{oktay2018attention}, has proven effective in feature extraction, enabling the network to capture the intricate dynamics inherent in ISWs.
In addition, the model incorporates multilayer perception (MLP) with differentiable physical constraints within its decoder modules, ensuring the dynamic consistency of the reconstructed flow fields. 
In response to the challenges posed by complex topography, the network employs a comprehensive approach that merges edge sampling optimization—enhancing the model's sensitivity to fluid dynamics near bottom boundaries—with numerical pre-processing of topographic data, thereby mitigating the broader impact of topographic variations on model propagation.

Three types of experiments are conducted to evaluate the performance of the proposed model:
comparative experiments against existing physical constraint models, assessing the quality of the high-resolution generated data;
ablation experiments, analyzing the contribution of individual modules, such as \BlockName, to the overall performance of \NetName;
and practical application tests using ISW field observation data from the South China Sea, evaluating the model's effectiveness in real-world scenarios.
The experimental results demonstrate that \NetName~outperforms existing models by achieving a $7.85\%$ improvement in PSNR and a $0.28\%$ improvement in SSIM compared to the benchmark method.
Detailed comparisons are provided in Table \ref{tab: different model on ISWs}.
Moreover, the high-resolution data generated by \NetName~effectively captures the intricate features of ISWs, as illustrated in Fig.\ref{fig: different model comparison}.
The ablation study highlights the critical role of the \BlockName~module in extracting high-frequency characteristics on spatial and temporal scales.
Its removal results in a noticeable decline in model performance, as shown in Table \ref{tab: ablation for modules}.
Lastly, \NetName~exhibits strong performance in reconstructing amplitude, velocity, and trailing waveform matching using ISW field observation data from the South China Sea.   
Notably, the trailing waveform matching error and amplitude reconstruction deviation are minimal, underscoring the model's applicability and robustness in real-world marine environments, as detailed in Fig.\ref{fig: satellite-observed visualization}.

\section{Related Work} \label{sec: Related Work}
Our work builds upon research on ocean numerical simulation models, hybrid physics-constrained deep learning approaches, STSR, and importance sampling.

\subsection{Ocean Numerical Simulations}
Ocean numerical simulations play a pivotal role in the community of physical oceanography, utilizing computational models to replicate and analyze the intricate dynamics of oceanic systems~\citep{marshall1997finite,yaocun1999numerical}. 
Many studies demonstrated the versatility of numerical models in simulating ocean circulation, mixing, and biogeochemical processes, offering foundational insights into ocean dynamics and climate interactions.
For instance, \citet{morey2020assessment} applied three distinct numerical general circulation models to reproduce the circulation in the Gulf of Mexico, showcasing the adaptability of numerical models to regional studies.
An alternative approach was proposed by~\citet{song2011comparisons}, who developed a time-domain numerical model to evaluate ISW-induced forces on marine structures, uncovering their potential threats to offshore safety.
According to~\citet{amores2022numerical}, two-dimensional ocean models are highly effective for simulating atmospheric Lamb waves, illustrating the connection between oceanic and atmospheric processes.
The approach was further developed by~\citet{wang2022numerical}, who investigated the spatio-temporal evolution of high-resolution internal tides in the Andaman Sea, providing deeper insights into their dynamics.

While traditional numerical simulation methods have advanced significantly, integrating low-resolution data with deep learning has emerged as a transformative approach. 
Recent developments, such as Fourier Neural Operators for PDE solutions~\citep{li2020fourier}, learning degradation models for blind super-resolution~\citep{luo2022learning}, and innovative convolutional modules like SPD-Conv~\citep{Sunkara2023No}, have demonstrated the potential of deep learning to drastically reduce the computational time required to generate high-resolution data. 
Such advancements represent a paradigm shift from conventional practices, enabling more efficient and scalable solutions in various fields.
To the best of our knowledge, the nonlinear characteristics of ISWs and the impact of complex bottom topography pose significant challenges to accurate representation~\citep{Xie2010a, Chen2023Seabed}. 
In this way, we utilized the MITgcm model to generate high-quality ISW datasets and developed \NetName~ by integrating Navier Stokes equations to process and analyze these datasets for enhanced ISW feature extraction and prediction.

\subsection{Hybrid Physics-constrained Deep Learning}
Hybrid physics-constrained deep learning integrates physical principles into neural networks, offering improved predictive accuracy while ensuring consistency with fundamental physical laws~\citep{ruthotto2020deep}, which has been widely applied in fields such as computational fluid dynamics, signal processing, and energy management~\citep{zhang2020learning}. 
It is necessary to employ domain decomposition techniques in deep learning research constrained by hybrid physics to enhance computational efficiency.
For instance, \citet{jagtap2020conservative} introduced cPINN, a framework that partitions computational domains into subdomains while maintaining flux continuity across their boundaries. 
Building on the framework of domain decomposition, \citet{jagtap2021extended} developed XPINN, which generalizes domain decomposition to arbitrary space-time partitions, enabling scalable modeling of diverse PDEs.

Hybrid physics-constrained deep learning has also been successfully applied in oceanographic studies. 
\citet{de2019deep} modeled sea surface temperature variations using an advection-diffusion equation, while \citet{chi2020fast} proposed MeshfreeFlowNet to address the super-resolution of Rayleigh–Bénard convection. 
More recently, \citet{wu2022physics} employed a hybrid LSTM-based model to improve the prediction of ISWs by embedding physical constraints into the neural network framework.
Building on these advances, \NetName~is a physics-constrained deep learning framework tailored for super-resolution oceanographic applications, with a particular emphasis on ISWs.

\subsection{Spatio-Temporal Super-Resolution (STSR)}
STSR builds upon innovations in video super-resolution, extending its capabilities to reconstruct high-resolution spatio-temporal data from low-resolution inputs~\citep{yue2023enhancing}. 
Video super-resolution enhances video quality by converting low-resolution videos into high-resolution counterparts, with applications in fields like entertainment, surveillance, and medical research~\citep{maity2023survey}. 
Most current video super-resolution methods rely on deep learning, using paired low-resolution and high-resolution datasets for training~\citep{Liu2022Video}. 
These models analyze low-resolution inputs to recover high-frequency details, producing superior-quality high-resolution outputs~\citep{yoon2015learning,kappeler2016video}. 
An efficient sub-pixel convolutional neural network for real-time video super-resolution was developed by \citet{shi2016real}, which significantly reduced computational costs by incorporating linear interpolation.
\citet{wen2022video} reported a spatio-temporal alignment network incorporating skip connections and attention mechanisms, optimizing video super-resolution by addressing spatio-temporal dynamics.

Building upon the principles of video super-resolution, STSR has emerged as a natural extension.
The approach is exemplified in the work of \citet{wang2021deep}, who utilized a deep residual network for super-resolution to improve the spatial resolution of daily precipitation and temperature datasets. 
Recent advancements in STSR include the physics-informed framework PhySR by \citet{REN2023PhySR}, which integrates temporal interpolation, convolutional recurrent refinement, and spatial reconstruction in latent space to improve simulation accuracy. 
Similarly, \citet{fanelli2024deep} developed a multi-scale deep learning model to enhance sea surface temperature resolution, capturing fine-scale features and reducing errors in the Mediterranean.
Deep learning techniques have also been applied to improve sea surface height resolution, as shown in the work of \citet{Martin2024Deep}, who revealed ocean eddy interactions and refined global flow assessments.
By embedding physical constraints tailored to spatio-temporal data, \NetName~delivers precise and efficient data upsampling, addressing challenges unique to STSR applications.

\subsection{Importance Sampling in Deep Learning}
Importance sampling has gained prominence in deep learning for its ability to streamline training and enhance model performance by focusing attention on critical data regions and addressing imbalanced sample challenges~\citep{chen2018fastgcn, wang2023octree}.
Recent studies have explored methods to allocate weights to training data regions based on specific features~\citep{shu2019push}. 
For instance, \citet{banerjee2021deterministic} refined mini-batch selection by minimizing the maximum mean discrepancy between selected and unselected samples, ensuring the sampled data better represent the overall distribution. 
Expanding on this approach, \citet{alain2015variance} proposed prioritizing training samples with high information content and optimal difficulty levels to accelerate convergence. 
Surveys such as that conducted by \citet{shu2023cmw} have developed a meta-model that adaptively learns explicit weighting schemes from data, effectively addressing challenges like class imbalance and label noise, with demonstrated transferability across tasks. 
Further advancing the field, \citet{lahire2023importance} advanced the theoretical understanding of importance sampling in deep learning by introducing metrics to evaluate sampling quality and exploring interactions with optimization methods to improve training efficiency and performance. 

These advancements underscore the growing utility of adaptive weighting and importance sampling in deep learning.
Despite these advancements, existing strategies often lack direct applicability to super-resolution tasks, especially in complex fluid dynamics. 
Resolving intricate terrain boundaries requires balanced attention across regions of varying importance, thus, \NetName~introduces a method that dynamically adjusts attention weights, meeting the unique demands of super-resolution tasks.

\section{Internal Solitary Wave Dataset} \label{sec: Internal Solitary Wave Dataset}
The datasets were categorized into three types: training datasets, evaluation datasets, and observation datasets. 
Specifically, the training and evaluation datasets, generated using the MITgcm model~\citep{marshall1997finite}, were designed to simulate ISW dynamics under controlled conditions, incorporating detailed tidal effects through optimized parameter settings.
To supplement these simulations, an observation dataset from field measurements in the South China Sea during spring 2001~\citep{ramp2004internal} was introduced, providing real-world benchmarks for model validation.
The observation dataset serves as a critical reference for model calibration and validation, ensuring that the numerical simulation results align with real-world ocean phenomena.

\begin{figure}[!h]
\centering
\includegraphics[width=0.8\textwidth]{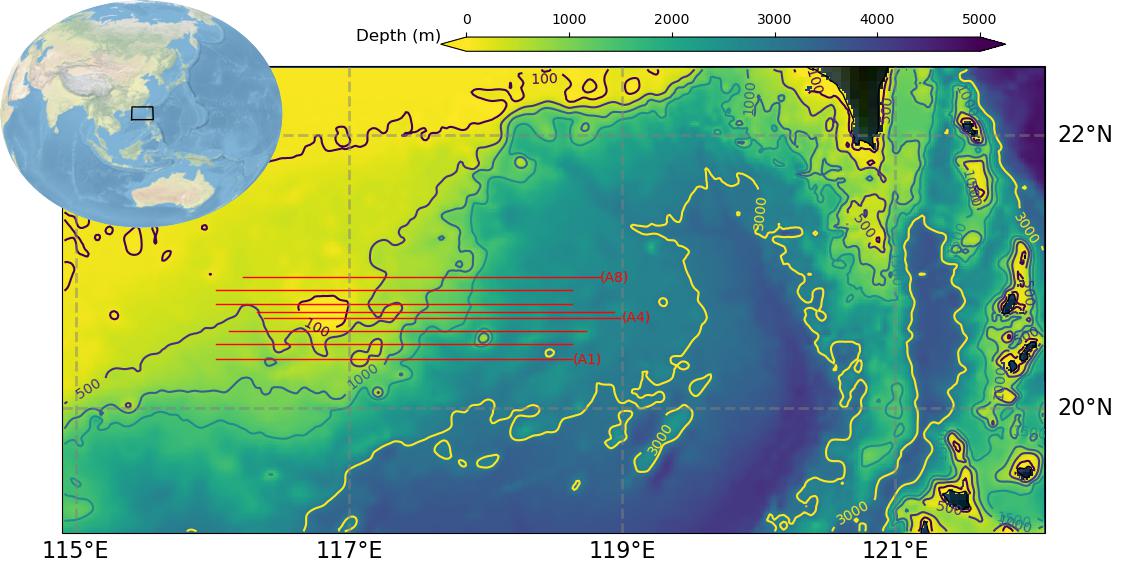}
\caption{Geographic distribution of the dataset capturing regional marine internal wave phenomena in the South China Sea. 
Subregion A(4) represents the validation set, while subregions A(1) to A(3) and A(5) to A(8) correspond to parts (a) to (g) of the test set, as detailed in Fig.\ref{fig: temperature profile of some datasets}.}
\label{fig: Slicing in the area}
\end{figure}

For the MITgcm simulations, it is a fully non-linear and non-hydrostatic three-dimensional numerical simulations with the area in a range of $115^\circ$E-$122.5^\circ$E and $17.7^\circ$N-$22.4^\circ$N as shown in Fig.\ref{fig: Slicing in the area}. The topography data is selected from the GEBCO2020 dataset~\citep{gebco2020}, while the eight tidal components are used at four boundaries to force the model, which consists of diurnal components ($K_1$, $O_1$, $P_1$, and $Q_1$) and semidiurnal components ($M_2$, $S_2$, $K_2$, and $N_2$) derived from the TPXO8-atlas dataset~\citep{Egbert1994topex}. The resolution is $500\,$m in the  propagation direction and $1000\,$m in the transverse direction. In the vertical direction, the non-uniform $147$ grids with a fine resolution in the upper and coarse resolution in the deep ocean are used. The validity of the model results and comparisons with observational data are shown in \citet{jia2024FMS}.     

\begin{figure}[!h]
\centering
\includegraphics[width=0.9\textwidth]{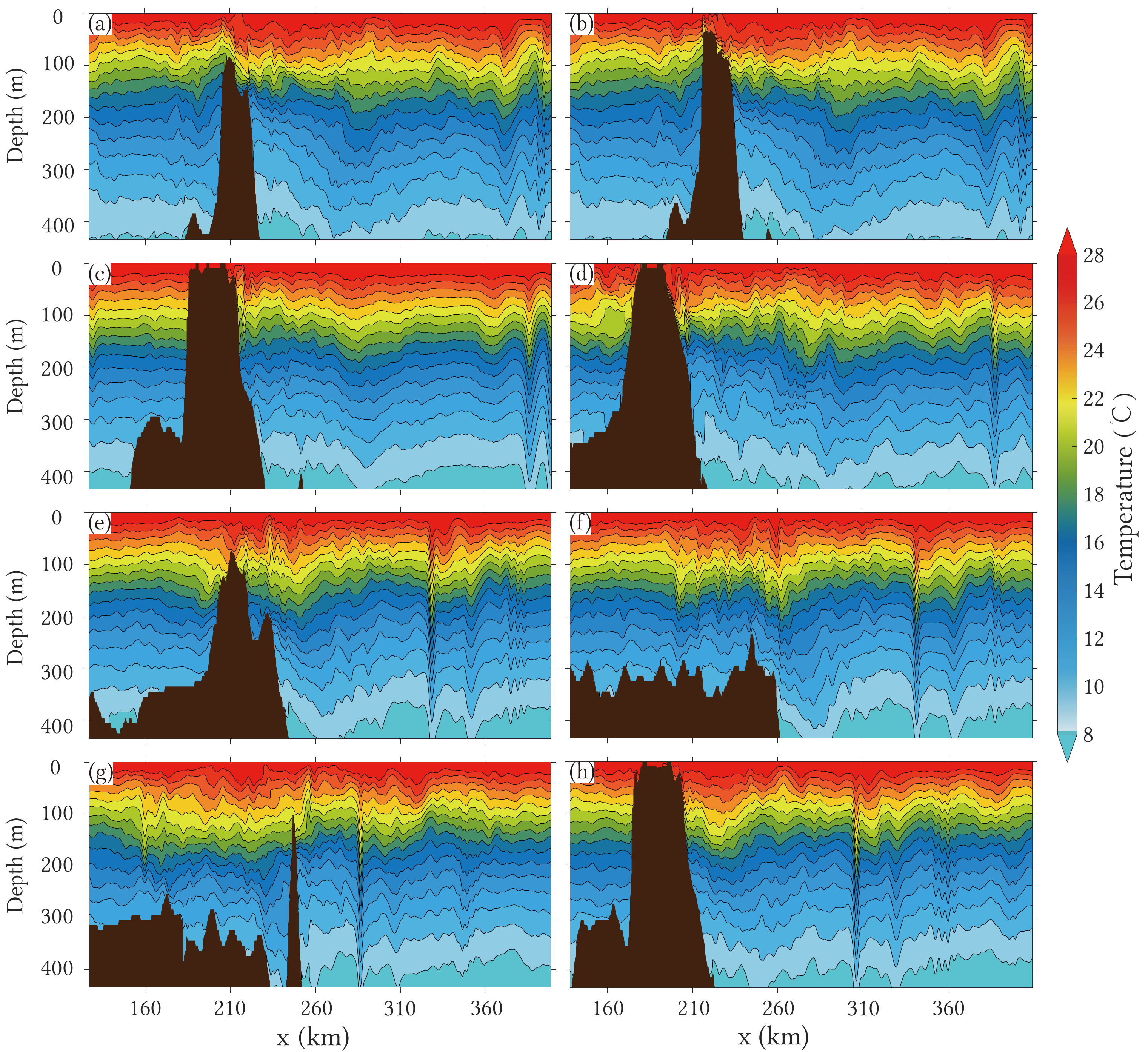}
\caption{The MITgcm model simulated temperature fluctuation induced by ISWs are shown along eight line sections at specific times, whose locations are shown in Fig.\ref{fig: Slicing in the area}.
Among them, datasets (a) through (g) are utilized for training the model, while dataset (h) is designated as the validation set. }
\label{fig: temperature profile of some datasets}
\end{figure}

This section provides a detailed discussion of the generation methods and characteristics of the ISW datasets. 
Key factors influencing the formation and dynamics of ISWs, such as topography, tidal forces, and their interactions, are analyzed. 
The oceanographic parameters, spatio-temporal scales, and other control variables related to PDEs are also elaborated in the context.
Fig.\ref{fig: temperature profile of some datasets} visualizes the waveform characteristics of different datasets at a specific moment, illustrating the distinct internal fluctuation patterns under varying conditions.

\subsection{Physical Constraints by the Navier-Stokes equations}
The physical characteristics of the generated data include key oceanographic parameters: temperature ($T$), salinity ($S$), velocity ($\vec{v}$), density ($\rho$), and pressure ($p$). 
These parameters determine the precision of the data generation process by governing the primitive Navier-Stokes equations ~\citep{brenner2008mathematical}. 
Additionally, incorporating physics-based constraints into neural networks can significantly enhance the effectiveness of model training~\citep{raissi2019physics}. Since, we consider the problems along every section line as in Fig.\ref{fig: Slicing in the area} (the dependence on $y$ is suppressed), then the two-dimensional Navier-Stokes equations, \textit{i.e.} $\vec{v}=(u,w)$ where $u$ and $w$ are particle velocities in the repsective $x$ and $z$ directions, can be written as follows:
\begin{equation}
\frac {\partial u} {\partial t} + u  \frac {\partial u} {\partial x} + w  \frac {\partial u} {\partial z} =  - \frac {1} {\rho} \frac {\partial p} {\partial x} + \nu_h \frac{\partial^2u}{\partial x^2} + \nu_z \frac{\partial^2u}{\partial z^2} \,,
\end{equation}
\begin{equation}
\frac {\partial w} {\partial t} + u  \frac {\partial w} {\partial x} + w  \frac {\partial w} {\partial z} =  - \frac {1} {\rho} \frac {\partial p} {\partial z} - g + \nu_h \frac{\partial^2w}{\partial x^2} + \nu_z \frac{\partial^2w}{\partial z^2} \,,
\end{equation}
\begin{equation}
\frac {\partial u} {\partial x} + \frac {\partial w} {\partial z} = 0 \,,
\end{equation}
\begin{equation}
\frac {\partial T} {\partial t} + u  \frac {\partial T} {\partial x} + w  \frac {\partial T} {\partial z} - K_{T} (\frac {\partial ^ 2 T} {\partial x ^ 2} + \frac {\partial ^ 2 T} {\partial z ^ 2}) = 0 \,,
\end{equation}
\begin{equation}
\frac {\partial S} {\partial t} + u \frac {\partial S} {\partial x} + w \frac {\partial S} {\partial z} - K_{s} (\frac {\partial ^ 2 S} {\partial x ^ 2} + \frac {\partial ^ 2 S} {\partial z ^ 2}) = 0 \,,
\end{equation}
\begin{equation} \label{eq:state}
\rho = F (T, S)\,,
\end{equation}
where $F$ in Eq.\eqref{eq:state} represent the Equation of State. This Navier-Stokes equation set is specifically applied in Section \ref{sec: Method}, where $K_T$ and $K_S$ are the temperature and salinity diffusion coefficients; $\nu_h$ and $\nu_z$ denote the lateral and vertical eddy viscosity, respectively; and $g$ is the acceleration due to gravity.
The values of the parameters $K_T$, $K_S$, $\nu_h$, and $\nu_z$ are as follows: $K_T = K_S = 0$, $\nu_h = 1 \times 10^{-3}\,$m$^2$/s and $\nu_z = 1 \times 10^{-6}\,$m$^2$/s. 

\section{Method} \label{sec: Method}
The architecture incorporates multiple \BlockName~units to efficiently capture high-frequency and complex features within the data. 
Additionally, \NetName~integrates a set of physical constraints closely aligned with ISW characteristics, significantly enhancing the model's performance for STSR tasks on ISW datasets.
To further improve the model's adaptability to complex terrain data, \NetName~introduces terrain region edge sampling optimization and numerical pre-processing techniques. 
These enhancements enable \NetName~to achieve more accurate training and predictions when processing ISW datasets with intricate topographic features of the seafloor.

\begin{figure}[!h]
\centering
\includegraphics[width=1.0\linewidth]{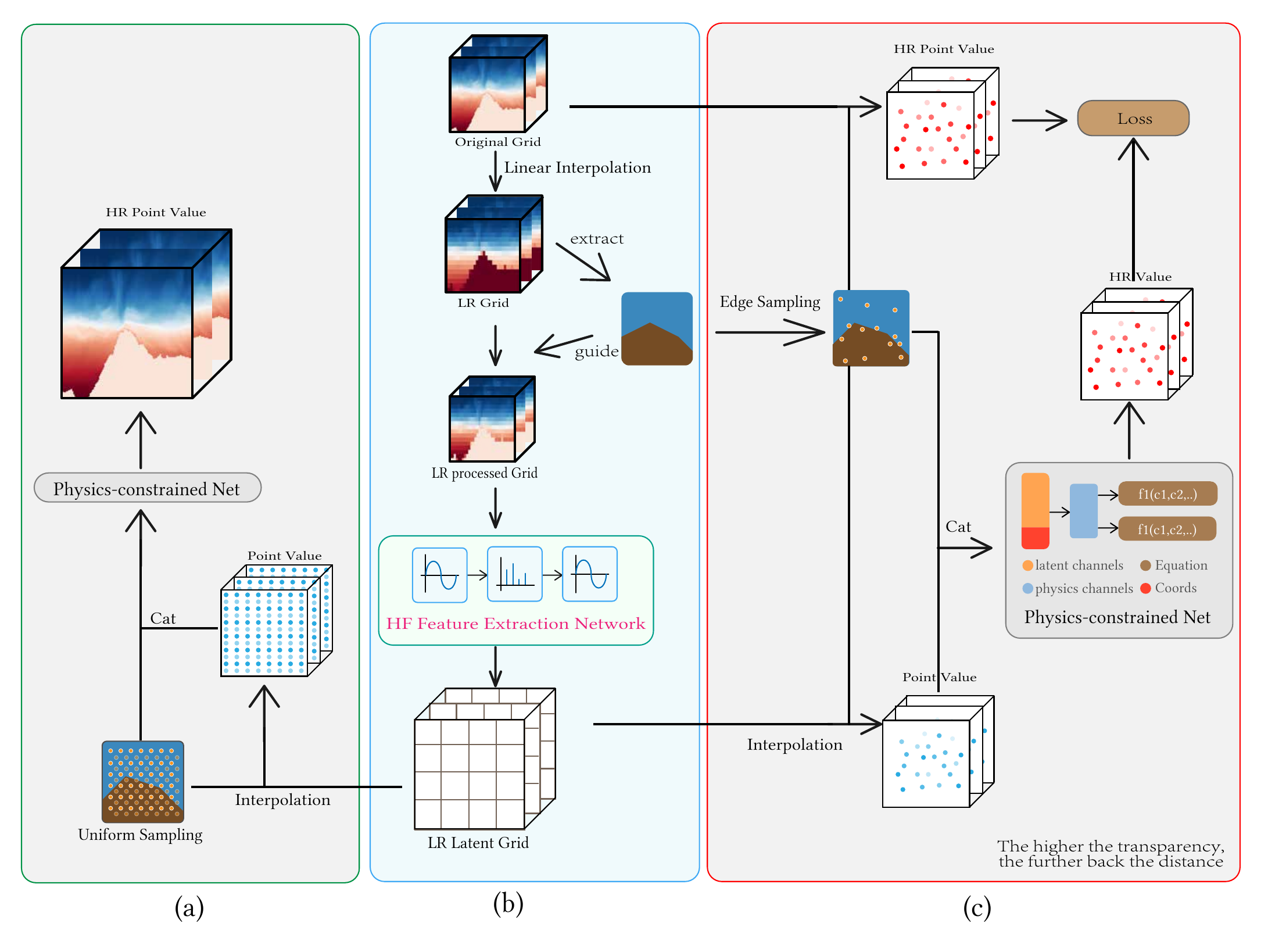}
\caption{
The architecture design and processing flow of \NetName~is illustrated in the figure, presenting the complete workflow for generating low-resolution grid data from raw input. 
First, topographic data is extracted and processed, combined with low-resolution implicit grid data, and coordinate point values are generated through interpolation. 
Next, the PCN refines the data and reconstructs high-resolution mesh features. 
Finally, the accuracy of generating high-definition images is assessed by comparing the results with ground truth data, and the model performance is further optimized through iterative loss calculations. 
For additional details, refer to Section \ref{sec: Method}.}
\label{fig: InWaveNet Structure}
\end{figure}
As illustrated in Fig.\ref{fig: InWaveNet Structure}, the structure of \NetName~consists of two key components: a high-frequency feature extraction network (HF-FEN) and a physics-constrained network (PCN). 
The HF-FEN is designed to enhance the realism of the generated results by extracting high-frequency features, while the PCN ensures physical consistency by incorporating domain-specific constraints. 
Fig.\ref{fig: InWaveNet Structure}b outlines the testing and validation process, which begins with generating implicit grids through interpolation operations. 
Low-resolution grid data, derived from the original data, is processed alongside extracted topographic features. 
The processed data is fed into the HF-FEN to produce low-resolution implicit grid data.

During the model optimization phase (Fig.\ref{fig: InWaveNet Structure}c), edge sampling points are utilized to interpolate low-resolution implicit grid data and terrain data, generating coordinate points. 
Combined with edge sampling information, these coordinate points are input to the PCN for prediction and refinement. 
The PCN is responsible for reconstructing high-resolution grid features while incorporating physical constraints that ensure the generated data remains consistent with ISW dynamics. 
Predicted coordinate values from the PCN are compared against ground truth values derived from interpolated raw data and edge coordinates, facilitating the computation of optimal loss functions for both the HF-FEN and PCN, which progressively refine model performance.

The verification and evaluation stage (Fig.\ref{fig: InWaveNet Structure}a) involves uniformly sampling topographic data and combining the sampled coordinate points with low-resolution implicit grid data. 
The combined data is then processed through the PCN to generate the final high-resolution results.
Transparency is inversely proportional to distance, with greater distances leading to lower transparency, thereby enhancing the qualitative effect of the rendering and ensuring alignment between the model output and actual observations.
Such an integrated approach ensures fidelity in high-resolution feature generation while improving the model's ability to manage complex terrain variations.
In summary, \NetName~provides a robust solution for ISW data STSR tasks by effectively capturing spatio-temporal and terrain-related features with high accuracy through advanced network architecture, physical constraints, and optimization techniques.

\subsection{Network Architecture}
\subsubsection{High-Frequency Feature Extraction Network (HF-FEN)}
\begin{figure}[!h]
\centering
\includegraphics[width=5.5in]{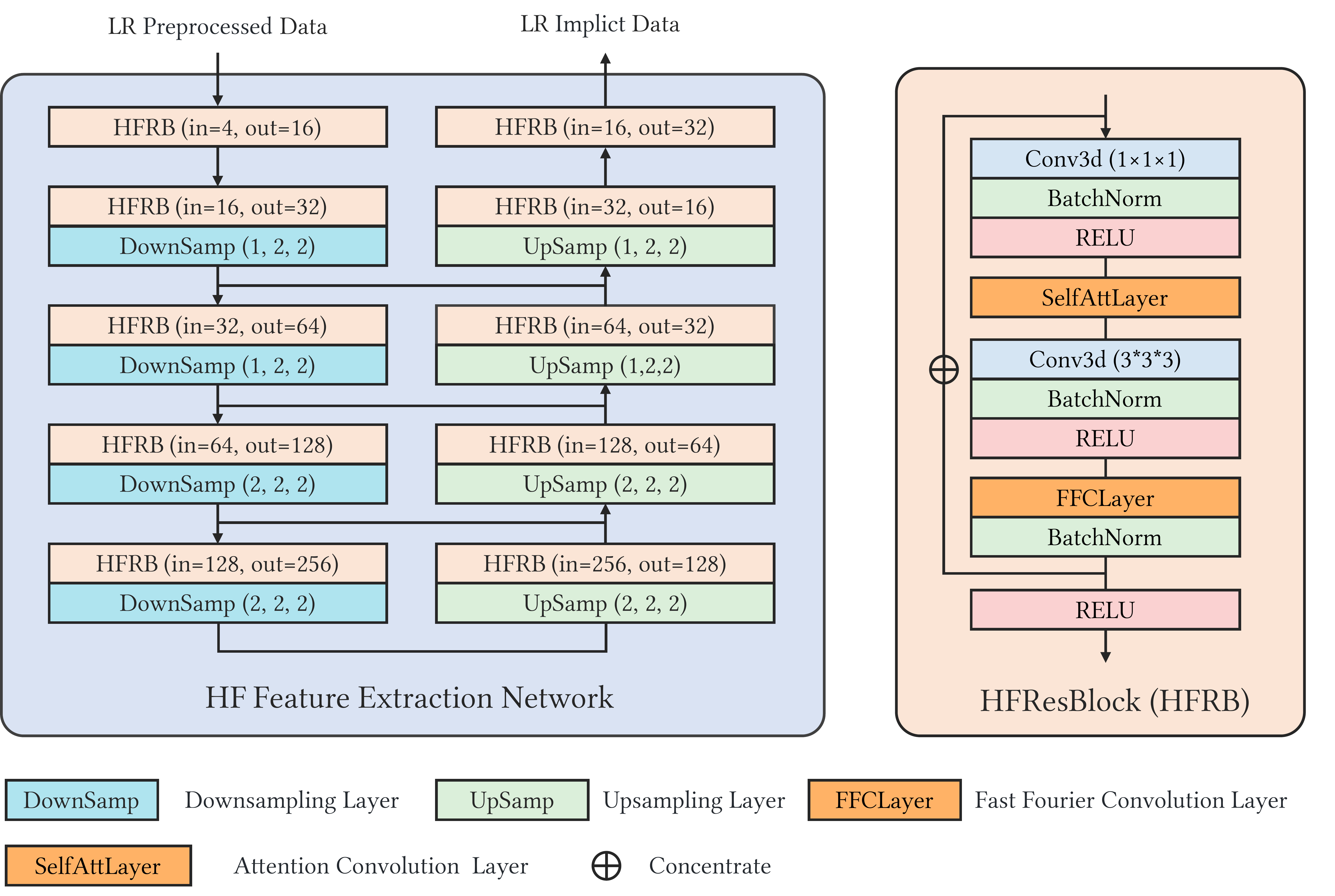}
\caption{
    The HF-FEN consists of multiple \BlockName s and exhibits a U-Net structure overall.
    }
\label{fig: HF Feature Extraction Network Structure}
\end{figure}

Within the \NetName~framework, we have integrated a novel feature extraction network, HF-FEN, which combines an attention module with an FFT, serving as an encoder for high-frequency features. 
As depicted in Fig.\ref{fig: HF Feature Extraction Network Structure}, HF-FEN is composed of multiple \BlockName~units and adopts an overall architecture inspired by U-Net~\citep{he2022swin, ronneberger2015u}. 
The design enhances the network's capability for feature extraction and representation.

The primary objective of the network is to establish a systematic mapping between physical variables and implicit features, represented as
$N_{FEN} (\cdot)$. 
This mapping enables the expression of the implicit feature grid as:
\begin{equation}
I_{LR}=N_{FEN} (T_{LR}),
\end{equation}
where $T_{LR}$ represents the low-resolution input data obtained from the numerical pre-processing of the terrain region. 
The output, $I_{LR} \in \mathbb{R}^{h \times w \times t \times c}$, contains $c$ channels corresponding to the number of original physical variables.

As depicted in Fig.\ref{fig: HF Feature Extraction Network Structure}, each attention and FFT-based residual block (HFRB) in HF-FEN consists of an attention mechanism module and an FFT layer to capture high-frequency features. The specific formula is:
\begin{equation}
X_{i+1} = 
\begin{cases}
    L_{HFRB (i)} (X_i), (i<=N/2), \\
    L_{cat} (L_{HFRB (i)} (X_i), (X_{N/2-i})), (2/N<i<N),
\end{cases}
\end{equation}
where $X_i \in \mathbb{R}^{ h_{i} \times w_{i} \times t_{i} \times c_{i} }$ represents the i-th input, ${L}_{HFRB (i)} (\cdot)$ denotes the i-th network substructure, ${L}_{cat} (\cdot)$ signifies the feature fusion function, and $N$ denotes the total number of HFRBs.

After obtaining the feature map from the FFT layer, the initial step involves extracting the real and imaginary parts. 
These parts are processed separately through convolution to effectively capture frequency-domain features. 
The processed real and imaginary parts are then merged, and the spatio-temporal representation is restored using the inverse Fourier transform. 
Based on this, the feature maps derived from the attention module are utilized to calculate spatio-temporal scores, which are subsequently multiplied with the convolutional feature maps. 
Such an operation enhances the network's weight allocation across both spatial and temporal dimensions, thereby optimizing its ability to represent high-frequency features.

The first half of the HF-FEN, as part of the overall architecture, employs a downsampling strategy to extract deep features, enabling the model to better capture correlations among grid features.
In the subsequent upsampling phase, the network reconstructs these features while integrating the feature maps from the earlier downsampling phase, ultimately producing an output with dimensions matching the input. 
As shown in Table \ref{table: the data shape of AttFftUnet3D}, the output dimension is $16 \times 16 \times 4 \times 32~(h \times w \times t \times c)$, where $c = 32$ indicates that each implicit feature contains 32 channels. 
During the inference process, low-resolution input data is fed into the HF-FEN, generating an implicit feature grid with the same spatio-temporal resolution as the input data.

\begin{table}[!h]
\caption{The design details of the HF-FEN.}
\label{table: the data shape of AttFftUnet3D}
\centering
\footnotesize
\begin{tabular}{lcc}
\hline
Input $(h \times w \times t \times c)$ & Operator & Channel  \\
\hline
$16 \times 16 \times 4 \times 4$ & HFRB & 16 \\
$16 \times 16 \times 4 \times 16$ & HFRB & 32 \\
$16 \times 16 \times 4 \times 32$ & DownSamp & - \\
$8 \times 8 \times 4 \times 32$ & HFRB & 64 \\
$8 \times 8 \times 4 \times 64$ & DownSamp & - \\
$4 \times 4 \times 4 \times 64$ & HFRB & 128 \\
$4 \times 4 \times 4 \times 128$ & DownSamp & - \\
$2 \times 2 \times 2 \times 128$ & HFRB & 256 \\
$2 \times 2 \times 2 \times 256$ & DownSamp & - \\
$1 \times 1 \times 1 \times 256$ & HFRB & 128 \\
$1 \times 1 \times 1 \times 128$ & UpSamp & - \\
$2 \times 2 \times 2 \times 128$ & HFRB & 64 \\
$2 \times 2 \times 2 \times 64$ & UpSamp & - \\
$4 \times 4 \times 4 \times 64$ & HFRB & 32 \\
$4 \times 4 \times 4 \times 32$ & UpSamp & - \\
$8 \times 8 \times 4 \times 32$ & HFRB & 16 \\
$8 \times 8 \times 4 \times 16$ & UpSamp & - \\
$16 \times 16 \times 4 \times 32$ & HFRB & 32 \\
$16 \times 16 \times 4 \times 32$ & - & - \\
\hline
\end{tabular}
\end{table}

\subsubsection{Physics-Constrained Network (PCN)}
Combined with the complex PCN method, the accuracy of flow simulation in computational fluid mechanics is significantly enhanced. 
The method not only delivers visually substantial super-resolution improvements but also adheres to strict scientific accuracy requirements.
The framework of the PCN method begins with a linear interpolation of point cloud coordinates ($p_i$) embedded in an implicit low-resolution dataset ($I_{LR}$). 
The interpolation of point cloud coordinates extracts potential point values, serving as the initial step in refining the dataset. 
The interpolated values are then combined with the original point cloud coordinates and processed through the physical information network to reconstruct high-resolution physical features ($G_{HR}$). 
The reconstruction process is governed by the network function $N_{PCN} (\cdot)$, which ultimately outputs high-resolution data at $p_i$.
The output of the high-resolution data is expressed as:
\begin{equation}
G_{HR} (p_i) = N_{PCN} (I_{LR} (p_i)) .
\end{equation}

The PCN consists of multiple fully connected layers and incorporates a residual structure.  
Embedding location information into the input features of each layer enhances the network's attention mechanism by leveraging positional information:
\begin{equation}
X_{i+1} =  L_{MLP (i)} (L_{cat} (X_i, p),
\end{equation}
where $X_i$ represents the i-th input, ${L}_{MLP (i)} (\cdot)$ denotes the i-th network substructure, and $p$ is point coordinate information.

Notably, PyTorch automatically computes the derivatives of features during regression loss calculation.  
The computed derivative values serve a dual purpose: it is utilized for backpropagation and can also be incorporated into the PDE for comparison.  
The PDE loss is defined based on these comparisons.  
The recovered physical features are further refined by incorporating location information and applying multiple physical constraint equations to calculate PDE losses.  
The process is formally defined as:
\begin{equation}
L_{PDE} = \sum{F_i (p_i, G_{HR} (p_i))},
\end{equation}
where $F_i$ represents a physical equation.
The total loss function is formulated as a weighted combination of the PDE loss and regression loss, expressed as:
\begin{equation}
L_{Tot} = \gamma L_{PDE} + (1-\gamma) (G_{HR} - T_{HR}),
\end{equation}
where $\gamma$ denotes the weight assigned to the PDE loss, and $T_{HR}$ represents the ground truth. 

\subsection{Optimization Techniques}
\subsubsection{Edge Sampling Optimization}
During the training phase, random sampling is employed to enhance the generalization ability of the network. 
However, the fluid dynamics at terrain edges introduce additional complexities, particularly in terms of reflection and collision phenomena. 
Random sampling can lead to overfitting in regions far from the edges, while edge regions may suffer from underfitting due to insufficient samples.

To overcome this question and accurately capture the features of terrain edge regions, we propose an edge sampling optimization method. 
The proposed approach significantly improves the performance of both edge regions and the overall network, as demonstrated in previous studies~\citep{katharopoulos2018not}. 
Specifically, the method prioritizes edge regions by shifting from global random sampling to preferential sampling points in edge regions.  
The sampling proportion for edge points is represented by the edge point coefficient $a$. 
During model training, $a$ is dynamically adjusted by tracking and comparing the average losses of edge points and random points, ensuring a balanced representation of both regions:
\begin{equation}
a = F_{a} (L_{e}, L_{r}),
\end{equation}
where $L_{e}$ and $L_{r}$ represent the average losses for edge and random points, respectively, and $F_{a}$ is a specific function designed for adjusting the edge sampling coefficient.
By tracking and balancing the losses of edge and non-edge points, the adaptive strategy enhances the network’s ability to learn both edge and non-edge features effectively

\begin{figure}[!h]
\centering
\includegraphics[width=0.8\textwidth]{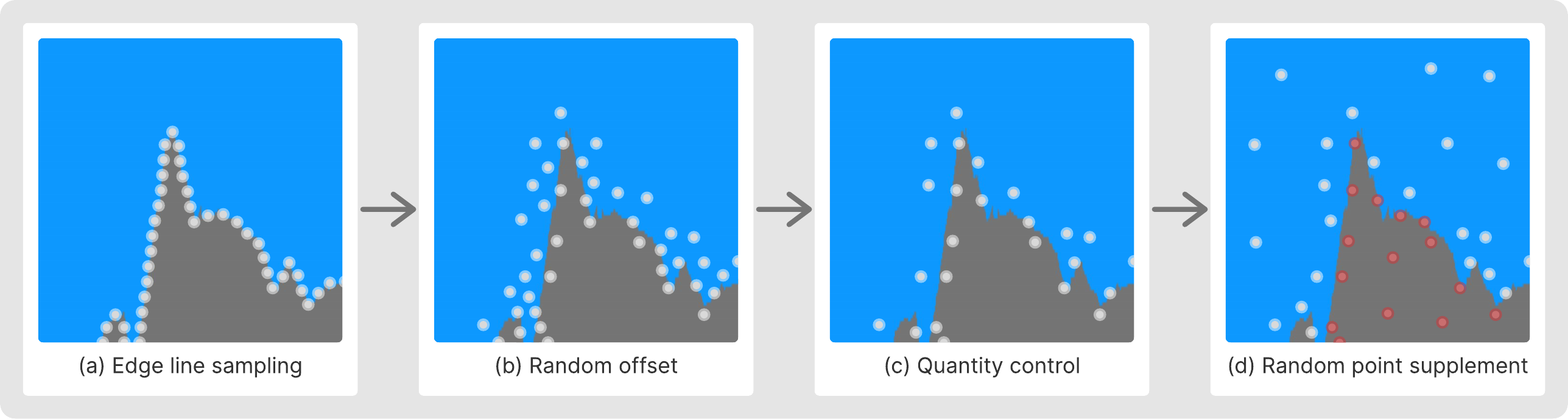}
\caption{
Edge sampling optimization. The process begins with the uniform selection of initial sampling points along the edge lines, proportional to their lengths.
These points are then subjected to small random shifts to preserve the edge features while maintaining their distribution near the edges. 
Subsequently, the shifted points are filtered to remove any that fall outside the grid area, ensuring the overall balance of the sampling proportion. 
Finally, the diversity of the dataset is further enhanced by randomly adding points from non-edge regions.
    }
\label{fig: The detailed steps of edge sample}
\end{figure}

Implementing the adaptive strategy requires several key steps in the edge sampling process, aimed at optimizing the representation of terrain edge features while ensuring overall dataset diversity.
As illustrated in Fig.\ref{fig: The detailed steps of edge sample}:
The edges are uniformly sampled, with the number of sampling points proportional to the edge length, expressed as $s \cdot L$, and the sampling point set is denoted as $\mathbf{p}_L$. 
Subsequently, the sampling points $\mathbf{p}_L$ are randomly shifted to distribute them near the edge line while preserving edge properties, represented as $\mathbf{p}_b = F_b (\mathbf{p}_L)$. 
The point set $\mathbf{p}_b$ is then filtered to remove points that may exceed the grid area due to offsets while maintaining the overall sampling proportion. 
The filtered point set is denoted as $\mathbf{p}_v = F_f (\mathbf{p}_b)$. 
Finally, to enhance the diversity of the point set, additional points are randomly selected from non-edge regions and combined with the filtered set $\mathbf{p}_v$ to form the final point set.
The edge sampling optimization addresses the imbalance caused by random sampling by prioritizing edge region features. 
It effectively enhances the network's comprehensive learning ability for both edge and non-edge regions.

\subsubsection{Numerical Pre-processing of Topography Regions}
In spatio-temporal grid data containing fluid and terrain components, terrain data values are typically fixed during generation and remain constant throughout training. 
These grid-based terrain data are provided to the model along with fluid data. 
However, significant differences between topographic and fluid values—particularly in cases of high topographic values—can result in inappropriate compression of the numerical range of the fluid data during normalization. 
Such compression degrades the accuracy of data representation and adversely affects the model's ability to capture fluid characteristics.

As a solution to this challenge, an integrated approach is employed prior to feeding spatio-temporal grid data into the model, where both fluid and topographic components undergo comprehensive pre-processing, coupled with edge sampling optimization to enhance the model's sensitivity to terrain-induced fluid dynamics.
A qualitative representation of the method is illustrated in Fig.\ref{fig: InWaveNet Structure}.
Specifically, \NetName~first identifies the fluid components in the feature grid data and independently calculates the average value of the fluid region for each feature:
\begin{equation}
\bar{V}_{\text{fluid}} = F_\text{aver} (x), \quad x \in \text{fluid},
\end{equation}
where $x$ represents the fluid component of the feature data. 
These computed average values are then assigned to the corresponding topography regions for each feature:
\begin{equation}
{V}_{\text{topo}} = \bar{V}_{\text{fluid}} ,
\end{equation}
where ${V}_{\text{topo}}$ denotes the adjusted numerical value of the topography. 
The adjustment ensures that, during subsequent standardization and normalization processes, the fluid component primarily determines the numerical range. 
Consequently, the proposed strategy effectively mitigates the adverse impact of topographic values on data processing.

\subsection{Implementation Details}
\textbf{Dataset parameters.}
The dataset utilized in the study is derived from numerical simulations of ISWs in the South China Sea, as detailed in~\citet{jia2024FMS}.   
Generated by the MITgcm project, the dataset formed a large-scale spatio-temporal grid with dimensions of $769 \times 147 \times 504 \times 1485 \ (t \times z \times y \times x)$.   
To facilitate a detailed analysis of ISW signatures, spatio-temporal slices exhibiting prominent ISW characteristics were carefully selected based on terrain features and observational data.   
These slices were subsequently adjusted to a standardized dimension of $256 \times 128 \times 512 \ (t \times z \times x)$.
To ensure data accuracy, the model's ability to replicate ISW dynamics was evaluated by comparing the numerical simulation results with satellite observation images.

\textbf{Dataset details.}
In the model workflow, one dataset is designated for evaluation, while the remainder is used for training.
A downsampling strategy is employed to address variations in interdimensional grid spacing. 
Specifically, the resolution is reduced by a factor of 4 along the $t$ and $x$ dimensions and by a factor of 8 along the $z$ dimension. 
During training, data blocks of size $16 \times 128 \times 128 \ (t \times z \times x)$ were randomly extracted to construct the dataset for a single epoch.  
These blocks were then downsampled to $4 \times 16 \times 32 \ (t \times z \times x)$ as low-resolution input data.

\textbf{Training details.}
The experiment was conducted on an Ubuntu 20.04 operating system with an Nvidia RTX 3090 GPU, Torch 1.13, Python 3.7, and CUDA 11.6. 
The Adam optimizer was used with a learning rate of $8 \times 10^{-4}$, and the model was trained for 100 epochs. 
Each epoch contained 8,000 randomly cropped data blocks, forming a feature grid dataset with 1,024 sample points. 
Due to memory constraints, the batch size was set to 6. 
The weight coefficient ${L}_{PDE}$, denoted as $\theta$, was set to $5 \times 10^{3}$.

\section{Results} \label{sec: Results and Analysis}
In this section, we provides a detailed visualization of the unique characteristics of ISWs, emphasizing their distinctive features and laying the groundwork for further analysis.
In Section \ref{sec: Comparison}, \NetName~is compared with other state-of-the-art visualization methods using quantitative metrics for a comprehensive evaluation. 
Additionally, Section \ref{sec: Ablation Study} is conducted to identify the key components contributing to the performance of the \NetName~method.
Following these analyses, Section \ref{sec: Validation Results on In-Situ Observational Data} conducts rigorous tests using observational ISW data to evaluate the generalization abilities of \NetName.
These tests validate its applicability and robustness in practical scenarios.

\begin{figure}[!h]
\centering
\includegraphics[width=0.7\textwidth]{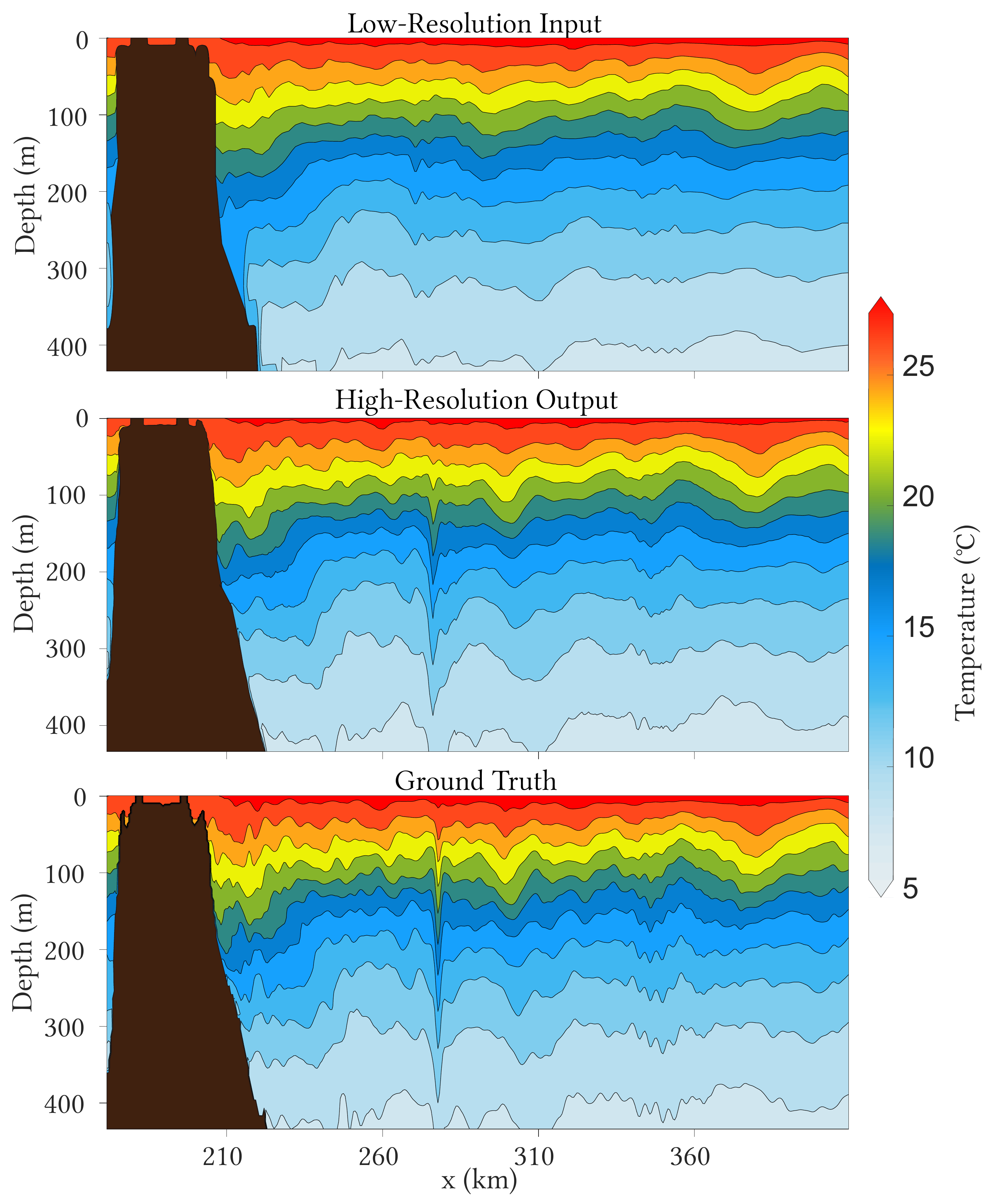}
\caption{
Visualization of the $x$-velocity at $t=56t$.
The diagram illustrates the process of converting low-resolution data (16$\times$128 grid) into high-resolution data (128$\times$512 grid) using \NetName.
Initially, \NetName~generates a low-resolution 16$\times$128 grid image by downsampling the original high-resolution temperature data.  
The downsampling step substantially reduces the visibility of ISWs and simplifies the data structure.  
Subsequently, \NetName~applies super-resolution reconstruction to restore the low-resolution data, producing results closely approximating the original high-resolution 128$\times$512 image.  
The super-resolution technique achieves a 32-fold resolution enhancement while significantly improving the detail and accuracy of the reconstructed data.}
\label{fig: slice heat u}
\end{figure}

\NetName~evaluates datasets through visualization using distribution and contour plots. 
The evaluation began by downsampling raw data to generate low-resolution datasets, which are then super-resolved into high-resolution datasets using \NetName.
As shown in Fig.\ref{fig: slice heat u}, \NetName's output achieves superior clarity and detail representation compared to the input image, demonstrating substantial enhancement. 
Comparative analysis indicates that \NetName~not only accurately reconstructs high-resolution data but also effectively captures intricate details absent in low-resolution input.

Specifically, at key locations around $x = 280km$, \NetName~revealed features that are undetectable with linear interpolation. 
The limitation of interpolation methods hampers their ability to identify key ISW characteristics. 
In contrast, \NetName~effectively captures these details, producing outputs with a high degree of accuracy comparable to ground truth data.
\subsection{Comparison}\label{sec: Comparison}
\subsubsection{Qualitative Evaluation}
\begin{figure}[!h]
\centering
\includegraphics[width=0.9\textwidth]{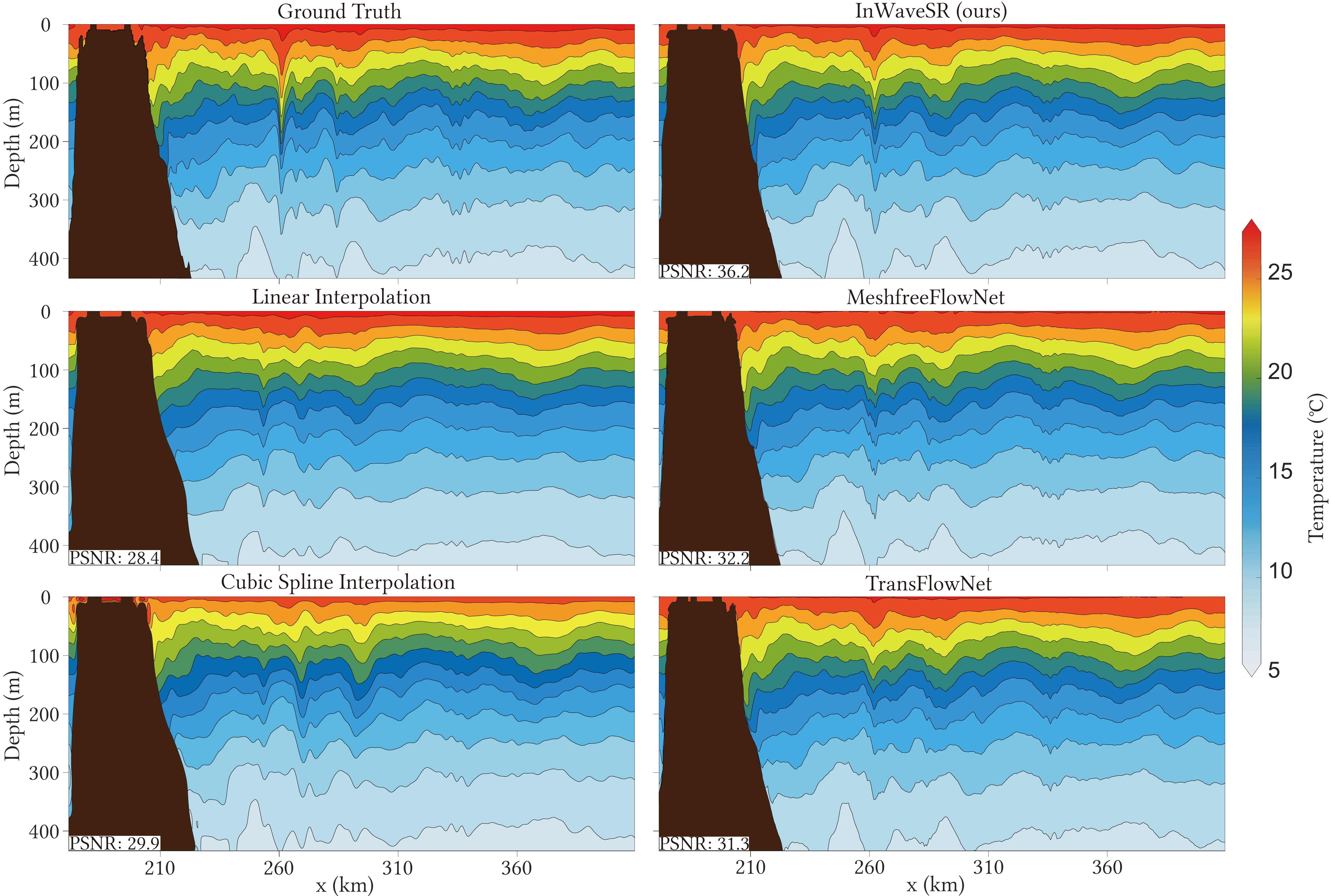}
\caption{
The brown area in the figure represents the terrain, with the numbers indicating the peak signal-to-noise ratio (PSNR) evaluation metric.  
A higher PSNR value corresponds to better image quality.  
The results demonstrate that the predictions generated by the \NetName~model not only achieve the closest visual alignment with the ground truth but also outperform other methods in terms of PSNR, fully validating the model's superior performance in salinity prediction.
}
\label{fig: different model comparison}
\end{figure}

Fig.\ref{fig: different model comparison} presents the visualization results of several methods: \NetName, Linear Interpolation, MeshfreeFlowNet, Cubic Spline Interpolation, and TransFlowNet, with GroundTruth serving as the reference data. 
These visualizations are evaluated using the PSNR, a metric where higher values indicate better image quality.
The figure clearly shows that \NetName~achieves the highest PSNR score of 32.2, demonstrating its superiority over the other methods. 
Such superiority is demonstrated by \NetName’s ability to accurately reproduce an internal wave with an amplitude of approximately 80 meters, closely aligning with the observed amplitude of 100 meters.
This accuracy underscores its ability to capture high-frequency ISW dynamics. 

ISW amplitude serves as a critical parameter for assessing internal wave energy and its influence on seabed topography, ocean currents, and ecosystems in oceanography.
\NetName’s accurate simulation of these dynamics highlights its potential for advancing ISW research.
The trailing wave plays a crucial role in the attenuation and dispersion of internal wave energy. 
\NetName~demonstrates outstanding performance in simulating the trailing wave, accurately reproducing the internal wave signal and its dynamic behavior, thereby advancing the study of internal wave.

\subsubsection{Quantitative Evaluation}
Our evaluation framework combines qualitative analysis with a range of quantitative metrics to comprehensively and systematically assess the performance of various approaches.

\textbf{Metrics.}
The quantitative evaluation includes traditional image quality metrics such as the structural similarity index (SSIM) and peak signal-to-noise ratio (PSNR)~\citep{hore2010image}. 
Additionally, physics-based metrics, including total kinetic energy error (KE-Error), assess the model’s ability to reconstruct ISW dynamics.
KE-Error quantifies the discrepancy between the predicted and actual waveforms, with lower values indicating better dynamic feature reproduction.
To further evaluate the model's ability to capture internal wave structures, we introduce the frequency-domain mean square error (FFT MSE) as a measure of frequency and phase accuracy. 
The FFT MSE is calculated using FFT to extract the frequency components of the physical grid features and then compute the mean square error between the predicted results and the ground truth.
Similar to KE-Error, a lower FFT MSE reflects better model performance in restoring complex waveform features.
\begin{table}[!h]
\caption{Quantitative comparisons of different methods on ISWs. 
The most favorable outcomes are highlighted in red, and the second in blue.}
\label{tab: different model on ISWs}
\centering
\footnotesize
\begin{tabular}{cccccc|c}
\hline
\multicolumn{2}{c}{\textbf{Metric}} 
& \begin{tabular}[c]{@{}c@{}}\textbf{Linear}\\\textbf{Interpolation}\end{tabular} 
&\begin{tabular}[c]{@{}c@{}}\textbf{Cubic Spline}\\\textbf{Interpolation}\end{tabular}
& \begin{tabular}[c]{@{}c@{}}\textbf{Meshfree-}\\\textbf{FlowNet}\end{tabular}
& \begin{tabular}[c]{@{}c@{}}\textbf{Trans-}\\\textbf{FlowNet}\end{tabular}
& \cellcolor{yellow!8}
\begin{tabular}[c]{@{}c@{}}\textbf{InWaveSR}\\\textbf{(ours)}\end{tabular}
  \\ 
\hline
\multirow{2}{*}{T} & PSNR$\uparrow$ & 28.4418 & 29.8560 & \textcolor{blue}{32.2308} & 31.2721 &\cellcolor{yellow!8}\textcolor{red}{36.1522}\\
&SSIM$\uparrow$ & 0.9436  & 0.9415  & \textcolor{red}{0.9784} & 0.9757
& \cellcolor{yellow!8}\textcolor{blue}{0.9781} \\ 

\multirow{2}{*}{S} & PSNR$\uparrow$& 25.0494& 26.6860& \textcolor{blue}{31.1333}& 30.0216& \cellcolor{yellow!8}\textcolor{red}{38.0843} \\
&SSIM$\uparrow$& 0.9025& 0.8959& \textcolor{blue}{0.9697}& 0.9657& \cellcolor{yellow!8}\textcolor{red}{0.9777}\\

\multirow{2}{*}{u} & PSNR$\uparrow$& 33.2437&33.1228& 37.8227 & \textcolor{blue}{37.9038}& \cellcolor{yellow!8}\textcolor{red}{37.9040} \\
&SSIM$\uparrow$& 0.8915& 0.8944 & 0.9581& \textcolor{blue}{0.9582}& \cellcolor{yellow!8}\textcolor{red}{0.9589} \\ 

\multirow{2}{*}{w} & PSNR$\uparrow$& 35.8201&35.1151& 38.9511
& \textcolor{red}{39.5579}& \cellcolor{yellow!8}\textcolor{blue}{39.0089}\\
&SSIM$\uparrow$ & 0.8965  & 0.8877  & 0.9513  & \textcolor{blue}{0.9533}  & \cellcolor{yellow!8}\textcolor{red}{0.9538} \\ 

\multirow{2}{*}{Avg.} & PSNR$\uparrow$& 30.6388& 31.1950& \textcolor{blue}{35.0345} & 34.6889 & \cellcolor{yellow!8}\textcolor{red}{37.7874}\\
&SSIM$\uparrow$ & 0.9085  & 0.9049  & \textcolor{blue}{0.9644}  & 0.9632 
& \cellcolor{yellow!8}\textcolor{red}{0.9671} \\ 
\hline
\multicolumn{2}{c}{KE-Error $\downarrow$} & 0.1777 & \textcolor{blue}{0.0498} & 0.0559 & \textcolor{red}{0.0472} 
& \cellcolor{yellow!8}0.0740 \\
\multicolumn{2}{c}{FFT MSE$\downarrow$ }  & 1.5533 & 1.7538 & \textcolor{blue}{1.0701} & 1.2915
& \cellcolor{yellow!8}\textcolor{red}{0.4322} \\
\hline
\end{tabular}
\end{table}
\textbf{Results.}
Table \ref{tab: different model on ISWs} summarizes the performance of various methods across multiple evaluation metrics. 
As shown in the table, our model outperforms others in two critical metrics: SSIM and PSNR, demonstrating the superiority of the proposed approach.
While KE-Error has some limitations as a metric, our method achieves the lowest FFT MSE, highlighting its significant advantage in capturing the frequency components of waveforms. 
Significant results highlight the high-frequency module's exceptional ability to reconstruct the dynamic characteristics of internal waves with high accuracy.

Noticeably, in Table \ref{tab: different model on ISWs} and Table \ref{tab: ablation for modules}, "T" represents temperature, "S" represents salinity, and "u" and "w" represent velocity components in the $x$ and $z$ directions, respectively. The term "Avg." refers to the global mean calculated previously. "KE-Error" denotes the total kinetic energy error, while "FFT MSE" corresponds to the mean square error computed in the frequency domain.

\subsection{Ablation Study} \label{sec: Ablation Study}
We conducted ablation studies on the ISW validation set to comprehensively evaluate the contribution of each module in the model. 
Specifically, the modules analyzed include \BlockName, which is designed to enhance high-frequency feature extraction; 
edge sampling optimization, which focuses on improving edge region accuracy; 
and numerical pre-processing of topography regions, which addresses the influence of terrain regions on fluid dynamics. 
The following sections discuss the specific setup and findings for each module in detail. 
The results of these studies are summarized in Table \ref{tab: ablation for modules}.

\begin{table}[!h]
\caption{
Ablation study results of \NetName~using the ISWs dataset.
The most favorable outcomes are highlighted in red, and the second in blue.}
\label{tab: ablation for modules}
\centering
\footnotesize
\begin{tabular}{ccccc|c}
\hline
\multicolumn{2}{c}{\textbf{Metric}} &\textbf{(w/o) ESO} & \textbf{(w/o) HF} & \textbf{(w/o) NPTR} & \cellcolor{yellow!10}\textbf{InWaveSR} \\ 
\hline
\multirow{2}{*}{T} & PSNR$\uparrow$& 36.1464 & \textcolor{blue}{36.1519}
& 35.9659 & \cellcolor{yellow!10}\textcolor{red}{36.1522}\\
&SSIM$\uparrow$ & 0.9772  & \textcolor{blue}{0.9780}  & 0.9779  & \cellcolor{yellow!10}\textcolor{red}{0.9781} \\
\multirow{2}{*}{S} & PSNR$\uparrow$& \textcolor{blue}{38.0841} & 38.0309
& 38.0808 & \cellcolor{yellow!10}\textcolor{red}{38.0843}\\
&SSIM$\uparrow$& \textcolor{red}{0.9783}  & 0.9769  & 0.9765  & \cellcolor{yellow!10}\textcolor{blue}{0.9777} \\
\multirow{2}{*}{u} & PSNR$\uparrow$ & \textcolor{blue}{37.9009}
& 37.8885 & 37.9001 & \cellcolor{yellow!10}\textcolor{red}{37.9040}\\
&SSIM$\uparrow$& 0.9583  & \textcolor{blue}{0.9586}  & 0.9585  & \cellcolor{yellow!10}\textcolor{red}{0.9589} \\
\multirow{2}{*}{w} & PSNR$\uparrow$& 39.0081 & \textcolor{red}{39.0717} & 39.0087 & \cellcolor{yellow!10}\textcolor{blue}{39.0089}\\
&SSIM$\uparrow$& \textcolor{blue}{0.9535}  & 0.9532  & 0.9533  & \cellcolor{yellow!10}\textcolor{red}{0.9538} \\
\multirow{2}{*}{Avg.} & PSNR$\uparrow$& 37.7849
 & \textcolor{blue}{37.7858} & 37.7389& \cellcolor{yellow!10}\textcolor{red}{37.7874}\\
&SSIM$\uparrow$& \textcolor{blue}{0.9668} & 0.9667  & 0.9666  & \cellcolor{yellow!10}\textcolor{red}{0.9671} \\
\hline
\multicolumn{2}{c}{KE-Error $\downarrow$}& 0.0784 & \textcolor{blue}{0.0765} & 0.0788 & \cellcolor{yellow!10}\textcolor{red}{0.0740} \\
\multicolumn{2}{c}{FFT MSE $\downarrow$ }& \textcolor{red}{0.4298} &0.4976& 0.4554 & \cellcolor{yellow!10}\textcolor{blue}{0.4322} \\
\hline
\end{tabular}
\end{table}
 
\textbf{\BlockName.}
We replaced \BlockName~with an equivalent number of convolutional layers to assess its importance. 
Comparative results indicate a significant performance decline in the absence of \BlockName, even when the number of convolutional layers remained constant.
Notably, the \BlockName-enhanced model demonstrated marked improvements in average PSNR and SSIM compared to models utilizing only convolutional layers. 
These findings validate the critical role of \BlockName~in extracting high-frequency features essential for capturing image details.

\textbf{Edge Sampling Optimization.}
We replaced edge sampling with random sampling and conducted a detailed comparative analysis to evaluate the impact of edge sampling optimization. 
Results revealed that models incorporating edge sampling optimization achieved significantly higher mean PSNR and SSIM values compared to those using random sampling.
These results highlight the advantages of edge sampling optimization in enhancing edge detail representation and overall image quality. 
By prioritizing edge regions, edge sampling optimization enables the model to more effectively capture fine features in edge areas, while also improving global image processing accuracy and visual quality.

\textbf{Numerical Pre-processing of Topography Regions.}
We omitted numerical pre-processing of topography regions and trained the model directly on raw data to evaluate its impact. 
Comparative analysis showed a noticeable decline in the model's ability to address the influence of terrain regions on fluid dynamics when numerical pre-processing of topography regions was excluded.
Models with numerical pre-processing of topography regions exhibited significant improvements in average PSNR and SSIM compared to those trained without pre-processing. 
These findings emphasize the pivotal role of numerical pre-processing of topography regions in optimizing the model's capacity to represent fluid characteristics and enhance overall image processing performance. 
By applying proper numerical pre-processing, the model achieves a more accurate understanding of complex fluid dynamic behaviors and simulates them more effectively.

\subsection{Validation Results on In-Situ Observational Data}
\label{sec: Validation Results on In-Situ Observational Data}
\begin{figure}[!h]
\centering
\includegraphics[width=0.5\textwidth]{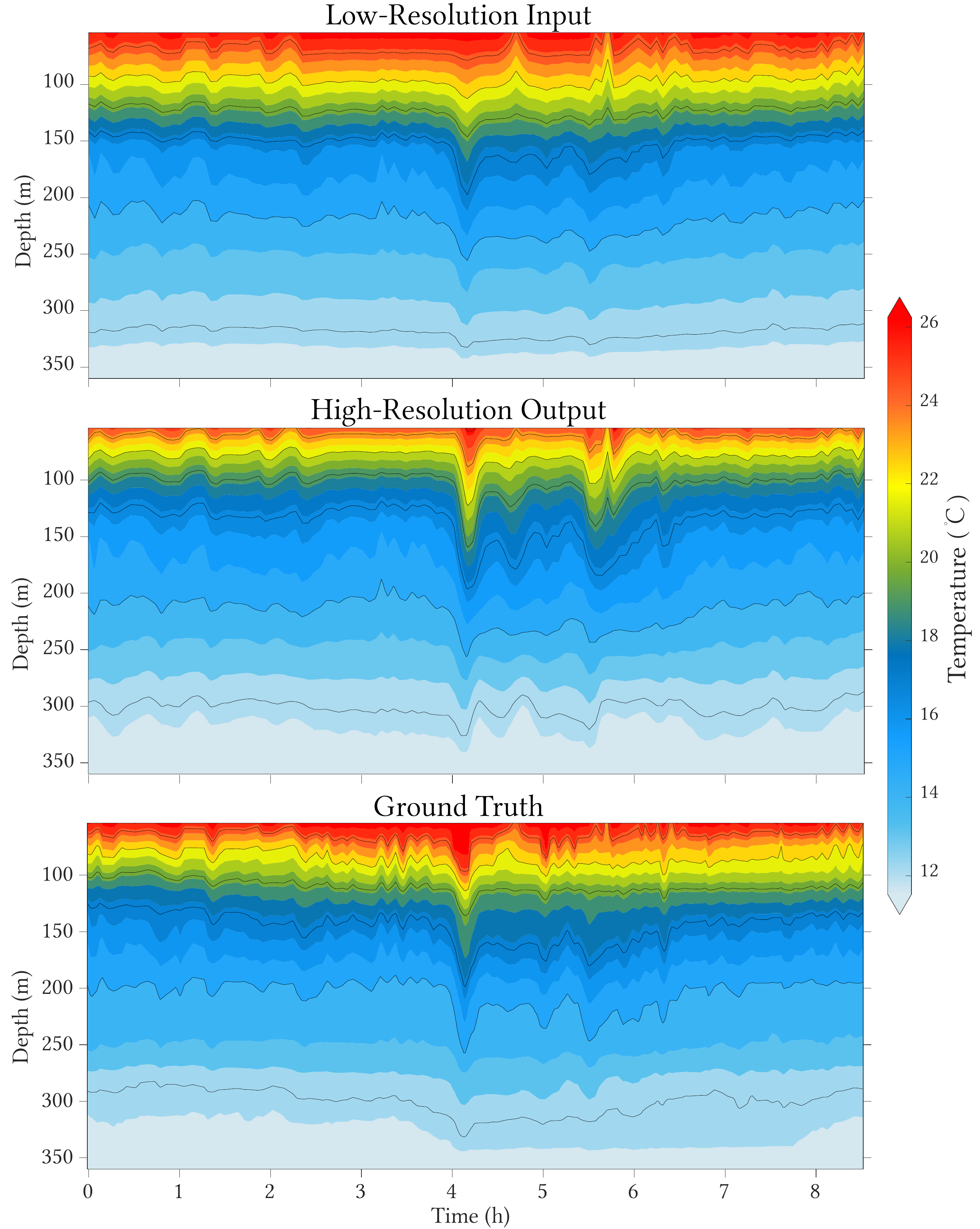}
\caption{
The graph, similar to Fig.\ref{fig: slice heat u}, utilizes the same super-resolution technology to transform low-resolution data from a $4 \times 128$ grid into high-resolution data on a $16 \times 512$ grid. 
The resolution enhancement process achieves a 16-fold resolution enhancement and successfully recovers details and structures that closely resemble the original high-resolution image.
}
\label{fig: satellite-observed visualization}
\end{figure}
To validate the generalization capability of \NetName, we conducted a series of experiments using field observational data, focusing on a critical study conducted in the South China Sea during the spring of 2001~\citep{ramp2004internal}. 
At first, the model was fine-tuned using a subset of the field observation data to enhance its adaptability to real-world observational data formats.

Following the fine-tuning, we evaluated the model using an independent subset of observations that were not included in the fine-tuning dataset. 
The evaluation results, illustrated in Fig.\ref{fig: satellite-observed visualization}, demonstrate \NetName's superior capability to reconstruct ISW waveforms from satellite observation data. 

\section{Conclusion} \label{sec: Conclusion}
In this study, we present \NetName, a novel ISW STSR model grounded in deep learning and augmented by physical constraints. 
The framework is specifically designed to achieve high-fidelity ISW STSR predictions by employing attention mechanisms and FFT techniques, enabling the extraction of high-frequency features from low-resolution input data.
By integrating physical equation analysis, \NetName~effectively quantifies the contribution of ISW dynamics to the overall system response. 
Integrating physical equations enhances the training process and ensures the model adheres to physical principles, thus improving prediction accuracy and reliability. 
Furthermore, \NetName~employs edge sampling optimization and numerical pre-processing of topography to enhance its capability in simulating and generating high-quality ISW datasets.
To evaluate the performance of \NetName, we conducted comprehensive numerical experiments and benchmarking against traditional models employing physical constraints and numerical simulation methods. 
The results demonstrate that \NetName~significantly outperforms conventional approaches, delivering superior accuracy and preserving intricate details in ISW STSR predictions.

However, certain limitations persist. 
A key challenge lies in the memory requirements for processing complex four-dimensional spatio-temporal datasets, which can strain computational resources. 
Future efforts will focus on refining the model architecture to enhance scalability and adaptability to diverse spatio-temporal grid dimensions.
Additionally, we aim to develop dynamic super-resolution capabilities within \NetName, enabling precise analysis and prediction of multiple feature combinations without necessitating model retraining.
Looking ahead, we intend to broaden the applicability of \NetName~to encompass a wider array of oceanographic phenomena and integrate additional ocean physical formulations. 
These advancements are expected to solidify \NetName~as a pivotal tool in marine science, offering a robust, efficient, and reliable framework for both research and practical applications.

\section*{CRediT authorship contribution statement}
\textbf{Xinjie Wang}: Writing – review and editing, Validation, Methodology, Funding acquisition, Formal analysis, Investigation, Conceptualization. 
\textbf{Zhongrui Li}: Writing – original draft, Visualization, Software, Data curation, Methodology. 
\textbf{Peng Han}: Writing – original draft, Software, Methodology. 
\textbf{Chunxin Yuan}: Writing – review and editing, Visualization, Validation, Data curation, Methodology, Funding acquisition, Investigation.
\textbf{Jiexin Xu}: Data curation, Investigation, Validation.
\textbf{Zhiqiang Wei}: Supervision,  Formal analysis.
\textbf{Jie Nie}: Funding acquisition, Supervision, Formal analysis.

\section*{Declaration of competing interest}
The authors declare that they have no known competing financial interests or personal relationships that could have appeared to influence the work reported in this paper.

\section*{Data availability} 
The datasets used in this study are available on GitHub. 
A link to access the datasets is provided in \url{https://github.com/spcrey/ISW_Dataset}. 

\section*{acknowledgments}
Chunxin Yuan was supported by the National Natural Science Foundation of China (\# 42476014), the Shandong Provincial Qingchuang Science and Technology Project (\# 2023KJ039), the Fundamental Research Funds for the Central Universities, China (\# 202442002, \# 202264007, \# 202265005), the State Key Laboratory of Tropical Oceanography, South China Sea Institute of Oceanology, Chinese Academy of Sciences, China (\# LTO2303). Xinjie Wang was supported by the National Key R\&D Program of China (\# 2022YFC2803805) and the Shandong Provincial Natural Science Foundation of China (\# ZR2024MF132), and Jie Nie was supported by the National Natural Science Foundation of China (\# U23A20320).

\bibliographystyle{elsarticle-harv}
\bibliography{reference}

\end{document}